\documentclass[12pt]{article}

\usepackage{amsmath}
\usepackage{amsfonts}

\begin{document}


\def\a{\alpha}
\def\b{\beta}
\def\c{\varepsilon}
\def\d{\delta}
\def\e{\epsilon}
\def\f{\phi}
\def\g{\gamma}
\def\h{\theta}
\def\k{\kappa}
\def\l{\lambda}
\def\m{\mu}
\def\n{\nu}
\def\p{\psi}
\def\q{\partial}
\def\r{\rho}
\def\s{\sigma}
\def\t{\tau}
\def\u{\upsilon}
\def\v{\varphi}
\def\w{\omega}
\def\x{\xi}
\def\y{\eta}
\def\z{\zeta}
\def\D{{\mit \Delta}}
\def\G{\Gamma}
\def\H{\Theta}
\def\L{\Lambda}
\def\F{\Phi}
\def\P{\Psi}
\def\S{\Sigma}
\def\V{\varPsi}
\newcommand{\EV}{ \,{\rm eV} }
\newcommand{\KEV}{ \,{\rm keV} }
\newcommand{\MEV}{ \,{\rm MeV} }
\newcommand{\GEV}{ \,{\rm GeV} }
\newcommand{\TEV}{ \,{\rm TeV} }

\def\o{\over}
\newcommand{\sla}[1]{#1 \llap{\, /}}
\newcommand{\beq}{\begin{eqnarray}}
\newcommand{\eeq}{\end{eqnarray}}
\newcommand{\gsim}{ \mathop{}_{\textstyle \sim}^{\textstyle >} }
\newcommand{\lsim}{ \mathop{}_{\textstyle \sim}^{\textstyle <} }
\newcommand{\vev}[1]{ \left\langle {#1} \right\rangle }
\newcommand{\bra}[1]{ \langle {#1} | }
\newcommand{\ket}[1]{ | {#1} \rangle }
\newcommand{\1}{\mbox{1}\hspace{-0.25em}\mbox{l}}


\baselineskip 0.7cm

\begin{titlepage}

\begin{flushright}
\end{flushright}

\vskip 1.35cm
\begin{center}
{\large \bf
A Note on Quasi-Riemannian Gravity with Higher Derivatives
}
\vskip 1.2cm
Izawa K.-I.
\vskip 0.4cm

{\it Yukawa Institute for Theoretical Physics, Kyoto University,\\
     Kyoto 606-8502, Japan}

{\it Institute for the Physics and Mathematics of the Universe, University of Tokyo,\\
     Chiba 277-8568, Japan}

\vskip 1.5cm

\abstract{
Quasi-Riemannian theories of gravity have smaller gauge groups
acting on the tangent spacetime than the full Lorentz group.
Among others, the spatial rotation group can be gauged to obtain
spacetime asymmetric gravity with general covariance.
We may introduce `spatial' higher-derivative interactions
exclusively to improve ultraviolet behavior with unitarity.
}
\end{center}
\end{titlepage}

\setcounter{page}{2}


General covariance deals with space and time on equal footing,
so that invariant interactions with space derivatives are accompanied
by those with time derivatives.
Hence higher derivative kinetic terms imply higher time derivative
ones, which results in unitarity-violating perturbative modes dubbed ghosts.

When the vielbein is introduced (to cope with fermions),
spacetime symmetry is usually maintained by imposing
local Lorentz symmetry. However, general covariance does not
necessarily require the full Lorentz group as the gauge group.
In fact, a subgroup of the Lorentz group can be taken as a gauge group
to construct a quasi-Riemannian theory of gravity
\cite{Wei,Gas}
with general covariance intact.

In particular, the rotation group of spatial dimensions
is a subgroup of the Lorentz group. If we adopt it as
the gauge group acting on the tangent spacetime,
the space and time may possibly be somehow independently treated
without symmetry among them in spite of general covariance.

Let $e^\mu_A$ denote a vielbein with $\mu=0,\cdots,d$ the spacetime
external index and $A=0,\cdots,d$ the tangent spacetime index
in $(d+1)$ spacetime dimensions.
The metric tensor $g^{\mu \nu}$ is defined as
$g^{\mu \nu}=\y^{AB}e^\mu_A e^\nu_B$ with the Minkowski metric
$\y^{AB}$, though we do not impose the full local Lorentz symmetry.
The index $A$ can be restricted to a spatial index $a=1,\cdots,d$
to obtain an internal spatial vector $e^\mu_a$
together with an internal scalar $e^\mu_0$,
both of which are external spacetime vectors with the index $\mu$.

Under the local rotation symmetry,
the kinetic terms with two time derivatives
of the corresponding quasi-Riemannian gravity
with dynamical variables $e^\mu_A$ can be obtained
\cite{Gas}
with several invariant terms
in addition to the Einstein-Hilbert term.
Higher derivative terms contain higher time derivatives
due to general covariance.

The spacetime derivative $\q_\mu$ is contracted with
the vielbein $e^\mu_A$ (except for the completely
antisymmetric invariant tensor) to form invariants
for general covariance.
We can restrict ourselves to use only $e^\mu_a$
to contract the derivative $\q_\mu$
such as $e^\mu_a \q_\mu$ for higher derivative terms.
On the other hand,
such invariant combinations as $e^\mu_0 \q_\mu$
may be confined in the two-derivative terms.

Although the resulting higher derivative terms still contain
the time derivative $e^0_a \q_0$,
they are harmless for perturbative unitarity
because of general covariance.
Namely, we can adopt an axial gauge $e^0_0=1$, $e^0_a=0$
to eliminate the problematic time derivatives (with the corresponding
`Gauss law constraints'). The remaining dynamical variables
consist of $e^i_0$ and $e^i_a$ with the index $i=1,\cdots,d$ the spatial
part in the spacetime index $\mu$, which need no negative
kinetic terms in contrast to the usual case
with the local Lorentz invariance.

Spatial higher derivative kinetic terms serve to improve
ultraviolet behavior in perturbative field theories
\cite{Ans}.
Renormalizable theories of gravity have already been
sought along such lines without general covariance
\cite{Hor}.
In the present quasi-Riemannian theories of gravity,
even under general covariance,
renormalizability may be achieved
with local dynamical degrees of freedom.
If so, the comparison with the Einstein gravity is of interest
at large length scales after renormalization.

\section*{Acknowledgements}

The author would like to thank T.~Kugo for valuable discussions.
This work was supported by the Grant-in-Aid for Yukawa International
Program for Quark-Hadron Sciences, the Grant-in-Aid
for the Global COE Program "The Next Generation of Physics,
Spun from Universality and Emergence", and
World Premier International Research Center Initiative
(WPI Initiative), MEXT, Japan.



\end{document}